\documentclass[prl,nopacs,twocolumn,preprintnumbers,notitlepage,amsmath,amssymb,superscriptaddress]{revtex4-1}

\usepackage{color}
\usepackage[pdftex]{graphicx}
\usepackage{pgffor}
\usepackage[]{pdfpages}

\makeatletter
\AtBeginDocument{\let\LS@rot\@undefined}
\makeatother

\definecolor{jerem}{rgb}{1, 0, 0}
\definecolor{jerem2}{rgb}{0, 0, 1}
\definecolor{olivier}{rgb}{0.125, 0.26, 0.07}

\newcommand{\der}[0]{\text{d}}

\usepackage{mathtools}
\usepackage{subcaption}

\usepackage{makecell}

\usepackage{enumerate}
\usepackage{mathtools}
\usepackage{stmaryrd}

\usepackage{enumitem}

\usepackage{epstopdf}

\usepackage{textcomp}
\usepackage{gensymb}
\def \equi#1{\mathrel{\mathop{\kern 0pt\sim}\limits_{#1}}}

\newcommand{\erf}[0]{\text{erf}}

\begin{document}
\title{Extreme Value Statistics of Jump Processes}
\author{J. Klinger}
\affiliation{Laboratoire de Physique Th\'eorique de la Mati\`ere Condens\'ee, CNRS/Sorbonne Université, 
 4 Place Jussieu, 75005 Paris, France}
\affiliation{Laboratoire Jean Perrin, CNRS/Sorbonne Université, 
 4 Place Jussieu, 75005 Paris, France}
\author{R. Voituriez}
\affiliation{Laboratoire de Physique Th\'eorique de la Mati\`ere Condens\'ee, CNRS/Sorbonne Université, 
 4 Place Jussieu, 75005 Paris, France}
\affiliation{Laboratoire Jean Perrin, CNRS/Sorbonne Université, 
 4 Place Jussieu, 75005 Paris, France}
 \author{O. B\'enichou}
\affiliation{Laboratoire de Physique Th\'eorique de la Mati\`ere Condens\'ee, CNRS/Sorbonne Université, 
 4 Place Jussieu, 75005 Paris, France}
 

\begin{abstract}
	

We investigate extreme value statistics (EVS) of general discrete time and continuous space symmetric jump processes. We first show that for unbounded jump processes, the semi-infinite propagator $G_0(x,n)$, defined as the probability for a particle issued from $0$ to be at position $x$ after $n$ steps whilst staying positive, is the key ingredient needed to derive a variety of joint distributions of extremes and times at which they are reached. Along with exact expressions, we extract novel universal asymptotic behaviors of such quantities. For bounded, semi-infinite jump processes killed upon first crossing of zero, we introduce the \textit{strip probability} $\mu_{0,\underline{x}}(n)$, defined as the probability that a particle issued from 0 remains positive and reaches its maximum $x$ on its $n^{\rm th}$ step exactly. We show that $\mu_{0,\underline{x}}(n)$ is the essential building block to address EVS of semi-infinite jump processes, and obtain exact expressions and universal asymptotic behaviors of various joint distributions.

\end{abstract}
\date{\today}

\maketitle

In a broad sense, \textit{extreme value} problems focus on the extrema of a set of random variables $(X_1,\dots,X_n)$. Determining the statistics of such extrema is of high practical interest to understand numerous physical systems driven by rare but extreme events. As an illustration, seismic risk evaluation \cite{Matthews:2002}, portfolio management \cite{Black:1973,Kou2003}  or understanding herd behavior \cite{Randon-Furling2009} are but a few examples of phenomena for which quantifying extreme value statistics (EVS) is key.  While EVS of sets of independent random variables have been studied early on \cite{Frechet:1927,Gumbel:1935}, leading to the renowned Gumbel-Frechet-Weibull universality classes for the distribution of the maximum of $n$ random variables, recent works have also focused on EVS of 
correlated random variables generated by single-particle trajectories, and more specifically, of continuous stochastic processes. Initiated by Paul Levy's \cite{Levy:1937,Levy:1939} derivation of the distribution of the running maximum $M(t)$ of a one dimensional Brownian particle $P(M(t)\leq M) = \erf\left(M/\sqrt{2t}\right)$,   
and the distribution of the time $t_m$ at which the running maximum is reached (also known as the arc-sine law)
\begin{equation}\label{eq:arcsinen}
	P(t_m=u|t)=\frac{1}{\pi \sqrt{u(t-u)}},
\end{equation}
\noindent a number of important results related to the EVS of one-dimensional Brownian dynamics have followed. In particular, joint distributions of extrema and times at which they are reached have been extensively studied for unbounded Brownian motions and Brownian bridges \cite{Majumdar:2008,Mori:2021,Mori:2020} as well as Brownian motions killed upon first passage to 0 \cite{Randon-Furling:2007,Klinger2022ab}.

Jump processes, which are discrete time and continuous space stochastic processes, constitute an alternative model to the continuous description of single particle dynamics.  
At each discrete time-step $n$, the particle performs a jump of length $\ell$ drawn from a distribution $p(\ell)$, whose Fourier Transform will be denoted $\tilde{p}(k)=\int_{-\infty}^\infty e^{ik\ell}p(\ell)\der \ell$.
Such processes are involved in various  contexts:  they constitute paradigmatic models of transport  in scattering media \cite{Baudouin:2014,Araujo:2021},  and  of self-propelled particles, living or artificial \cite{Romanczuk:2012fk,Tejedor:2012ly,Levernier:2021aa,Meyer:2021tx,Mori2020a}.
Most importantly, jump processes are particularly suited to describe inherently discrete empirical time series, where continuous stochastic models fail to capture discretization effects. As an illustration, the experimentally measured transmission probability of photons through 3D slabs \cite{Baudouin:2014,Araujo:2021} has been shown to be equivalent to the splitting probability $\pi_{0,\underline{x}}(0)$ that a jump process originated from $0$ crosses $x$ before 0 \cite{Klinger2022}. 
Accurately characterizing the EVS of jump processes is thus essential to quantitatively describe associated empirical measurements.

For symmetric jump processes considered hereafter, general EVS results are scarce, and  primarily focused on two types of observables.
First, the distributions of the time $n_m$ at which the maximum is reached \cite{sparre,feller-vol-2} and of successive record-breaking times \cite{Majumdar:2010r} have been shown to be independent of $p(\ell)$, and computed exactly.
Second, the asymptotic distribution of the running maximum $M_n$ has been studied  in the scaling limit and can be found in Darling \cite{Darling:1956} (see SM for details). 
Note however that  the specific behavior of $M_n$ stemming from the discrete nature of jump processes has only been characterized at the level of the  
the expected value of $M_n$, which has been investigated for processes with $\int \ell p(\ell)\der \ell<\infty$. In particular, the leading order large $n$ behavior of $\mathbb{E}(M_n)$ has been shown \cite{Comtet:2005,Mounaix:2018,DeBruyne:2021} to only depend on the tails of $p(\ell)$, equivalently described by the small $k$ expansion of $\tilde{p}(k)$
\begin{equation}\label{eq:small_fourier}
	\tilde{p}(k)\underset{k\to0}{=}1-(a_\mu |k|)^\mu+o(|k|^\mu).
\end{equation}
\noindent Here, the Levy index $\mu\in]0,2]$ describes the large $\ell$ behavior of $p(\ell)$, and $a_\mu$ is the characteristic lengthscale of the jump process. Importantly, when $\mu<2$, the jump process is dubbed heavy-tailed, and the jump distribution decays algebraically: $p(\ell)\propto \ell^{-(1+\mu)}$. 

\medskip

\begin{figure}[h!]
	\begin{subfigure}[T]{0.48\columnwidth}
	\centering
	\begin{picture}(100,90)
		\put(0,0){\includegraphics[]{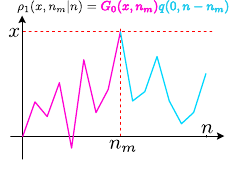}}
		\put(-5,50){(a)}
	\end{picture}
\end{subfigure}
	\begin{subfigure}[T]{0.48\columnwidth}
	\centering
	\begin{picture}(100,90)
	\put(0,0){\includegraphics[]{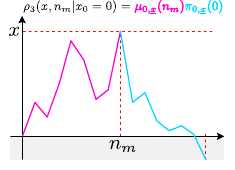}}
	\put(-5,50){(b)}
	\end{picture}
\end{subfigure}
\caption{(a) Sample trajectory contributing to the joint distribution $\rho_1$ of the maximum $x$ and time at which it is reached $n_m$ for an unbounded $n$-step long process. Since the survival probability $q(0,n)$ can be obtained from $G_0(x,n)$, computing $\rho_1$ reduces to evaluating the semi-infinite propagator.
(b) Sample trajectory contributing to the joint distribution $\rho_3$ of the maximum $x$ and time at which it is reached $n_m$ for a semi-infinite process. The derivation of $\rho_3$ requires the knowledge of the strip probability $\mu_{0,\underline{x}}(n)$.}
\label{fig:schem}
\end{figure}

\textit{General outline.} In the following, we develop a general framework to systematically analyze EVS of symmetric jump processes originating from 0. We show that computing joint distributions of EVS observables reduces to the evaluation of two key quantities: the semi-infinite propagator $G_0(x,n)$, defined as the probability that the particle 
remains positive and reaches $x$ on its $n^\text{th}$ step, and the \textit{strip probability} $\mu_{0,\underline{x}}(n)$, defined as the probability that the particle 
remains positive and reaches its maximum $x$ on its $n^\text{th}$ step exactly. The main result of this letter is the derivation of an exact expression of  $\mu_{0,\underline{x}}(n)$, and the analysis of its large $x$ and $n$ limit for general jump processes. In turn, we obtain exact expressions for a variety of new joint distributions of EVS observables, from which we uncover universal asymptotic behaviors.
These joint distributions, summarized in table \ref{tab:summary}, span both unbounded jump processes with deterministic number of steps $n$ (figure \ref{fig:schem}(a)), and bounded, semi-infinite jump processes killed upon first crossing of 0 (figure \ref{fig:schem}(b)), for which the discrete nature of the dynamics plays a crucial role.
 While the main text focuses exclusively on jump processes with continuous $p(\ell)$ originating from zero, our framework is easily extended to non-zero initial conditions, as well as lattice random walks (see SM).


\medskip

{\it EVS of unbounded jump processes.} In this section, we focus on general $n$-step long unbounded jump processes issued from 0.
By means of introduction, we consider the distribution $\mu(x|n)$ of the running maximum. To highlight the significant role of the semi-infinite propagator in EVS computations, we first recall a few important known results (equations \eqref{eq:max_distrib_1} to \eqref{eq:semiinf_propagator_2} and \eqref{eq:max_distrib_3}).
Defining the survival probability $q(x_0,n)$ that a particle issued from $x_0$ remains positive during its first $n$ steps, it is easily seen that \cite{Majumdar:2010r}
\begin{equation}\label{eq:max_distrib_1}
	\mu(x|n)=\frac{\der}{\der x}q(x,n).
\end{equation}
\noindent In turn, the survival probability is given by $q(x_0,n)=\int_{0}^{\infty}G(x,n|x_0)\der x$, where the semi-infinite propagator $G(x,n|x_0)$, defined as the probability that the $n$-step long trajectory issued from $x_0$ stays positive and is at position $x$ after $n$ steps, is known \cite{Ivanov1994,Majumdar:2010r}, and reads in Laplace and generating function space:
\begin{equation}\label{eq:semiinf_propagator}
	\begin{split}
	\sum_{n=0}^{\infty}\xi^n\left[\int_0^\infty  \int_0^\infty e^{-s_1x+s_2 x_0} G(x,n|x_0) \der x \der x_0\right]\\
	=\frac{\Tilde{G}_0(s_1,\xi)\Tilde{G}_0(s_2,\xi)}{s_1+s_2}	
		\end{split}
\end{equation}
\noindent where $\Tilde{G}_0(s,\xi)=\sum_{n=0}^{\infty}\xi^n\left[\int_{0}^{\infty}e^{-sx}G_0(x,n)\der x\right]$ is the Laplace transform of
$G_0(x,n)\equiv G(x,n|0)$, and is given in terms of $\tilde{p}(k)$ only by the Pollazceck-Spitzer formula \cite{Spitzer:1956,Pollacezk:1952}
\begin{equation}\label{eq:semiinf_propagator_2}
\Tilde{G}_0(s,\xi)= \text{exp}\left[-\frac{s}{2\pi}\int_{-\infty}^{\infty}\frac{ \ln\left[1-\xi \tilde{p}(k)\right]}{s^2+k^2} \der k \right].
\end{equation}
\noindent While equation \eqref{eq:max_distrib_1} is exact, it is clear from equation \eqref{eq:semiinf_propagator_2} that explicit expressions for the distribution of the running maximum can only be obtained for specific jump distributions.  
For instance, in the case of the exponential jump process $p(\ell)= 2^{-1}e^{-|\ell|}$, the semi infinite propagator can be found in \cite{Majumdar2017}, from which we explicitly derive the generating function of $\mu(x,n)$:
\begin{equation}\label{eq:max_laplace}
\sum_{n=0}^{\infty} \xi^n \mu(x|n)=\frac{(1-\xi-\sqrt{1-\xi})e^{-x\sqrt{1-\xi}}}{\xi-1}.
\end{equation}
\noindent We emphasize that for jump processes for which the semi-infinite propagator cannot be obtained explicitly, equations \eqref{eq:max_distrib_1} and \eqref{eq:semiinf_propagator_2} still allow for the asymptotic analysis of $\mu(x|n)$, which depends only on the Levy index $\mu$ and lengthscale $a_\mu$.
Defining $n_x\equiv(x/a_\mu)^\mu$ as the typical number of steps needed to cover a distance $x$, we first consider the large $n$ and $x$ scaling limit with $\tau\equiv n/n_x$ fixed. In this limit, jump processes are known to converge to Brownian motion \cite{Redner:2001a} with $D=a_2^2$ when $\mu=2$, and symmetric $\alpha$-stable processes \cite{Kyprianou:2006} when $\mu<2$. In turn, the limit distribution of the running maximum is given by Darling's result \cite{Darling:1956} (see SM for explicit expressions).
In the alternative limit regime $1\ll n \ll n_x$, the behavior of $\mu(x|n)$ for processes with $\mu=2$ depends on the details of $p(\ell)$. However, for heavy-tailed processes the distribution of $M_n$ becomes universal, and is readily obtained by extracting the leading order behavior of $G_0(x,n)$ from equation \eqref{eq:semiinf_propagator_2} (see SM), yielding:
\begin{equation}\label{eq:max_distrib_3}
	\mu(x|n)\underset{\substack{1\ll n\ll n_x}}{\sim}\frac{\mu n}{\pi}\sin\left(\frac{\pi \mu}{2}\right)\Gamma(\mu)\left[\frac{a_\mu}{x}\right]^\mu \frac{1}{x}.
\end{equation}
\noindent Importantly, the linear dependence of $\mu(x|n)$ admits a single big jump physical interpretation \cite{Vezzani:2019}: the particle has exactly $n$ trials to perform a very large jump bringing it close to $x$. Of note, the algebraic behavior \eqref{eq:max_distrib_3} can also be recovered by analyzing the asymptotic behavior of the maximum distribution of $\alpha$-stable processes \cite{Bingham:1973}. 
The semi-infinite propagator is thus an essential tool to derive exact and asymptotic expressions of $\mu(x|n)$. More generally, we claim that it is the necessary and sufficient building block to analyze arbitrary joint space and time EVS distributions, which we illustrate by computing two important quantities.

%
%
%

 We first determine the classical joint distribution $\rho_1(x,n_m|n)$ of the maximum $x$ and time $n_m$ at which it is reached, 
which, so far, has only been derived exactly for continuous processes.
By splitting the Markovian trajectory at $n_m$ (see figure \ref{fig:schem}(a)), and identifying the probabilistic weights of the first and second independent parts, the joint distribution is given by
\begin{equation}\label{eq:rho1_1}
	\rho_1(x,n_m|n)=G_0(x,n_m)q(0,n-n_m).
\end{equation}
\noindent When $\mu=2$, the asymptotic behavior of $\rho_1(x,n_m|n)$ is simply given by the corresponding Brownian result obtained in \cite{Borodin1996}. When $\mu<2$, no $\alpha$-stable limit result exists; in turn, we analyze the large $x$, $n_m$ and $n$ limit of equation (8), and uncover emerging universal behavior of $\rho_1(x,n_m|n)$ which depends only on $a_\mu$ and $\mu$: 
 \begin{equation}\label{eq:rho2_3}
	\hspace{-10pt}		\rho_1(x,n_m|n)\underset{\substack{n_x/n\gg 1\\ n/n_m\gg 1 }}{\sim}\frac{1}{\pi}\sqrt{\frac{n_m}{n-n_m}}\frac{2\mu}{\pi}\sin\left(\frac{\pi \mu}{2}\right)\Gamma(\mu)\left[\frac{a_\mu}{x}\right]^\mu \frac{1}{x}.
\end{equation}
\noindent In fact, our framework permits a more detailed characterization of space and time statistics, as we show by  providing the refined multivariate distribution $\rho_2(x,n_m,x_f|n)$ of the maximum $x$, time $n_m$ at which it is reached, and last position $x_f$ of the particle in terms of $G_0(x,n)$ only:
\begin{equation}\label{eq:ch6_JL_tri}
	\rho_2(x,n_m,x_f|n)=G_0(x,n_m)G_0(x-x_f,n-n_m).
\end{equation}
\noindent The asymptotic behavior of  $\rho_2(x,n_m,x_f|n)$ can be readily obtained for any $\mu$ from this general expression as is shown in SM.
Finally, we have shown that studying EVS of unbounded jump processes reduces to the evaluation of a single essential quantity: the semi-infinite propagator $G_0(x,n)$. In the following, we extend these results to the  case of bounded, semi-infinite jump processes.



{\it EVS of semi-infinite jump processes.} We consider jump processes killed upon crossing 0 for the first time, and hereafter choose $x_0=0$, although all our results are easily adapted to non-zero initial conditions (see SM). Note that EVS are properly defined for semi-infinite jump processes starting from zero, in striking contrast to corresponding EVS of continuous processes killed upon first passage to 0, which, by definition, vanish as $x_0\to 0$. Following the unbounded case, we first compute the distribution $\mu_{\underline{0}}(x|0)$ of the maximum $M_{\underline{0}}$ reached before crossing 0. Recalling the definition of the splitting probability $\pi_{0,\underline{x}}(0)$, it is clearly seen that the cumulative distribution of $M_{\underline{0}}$ satisfies $\int_{0}^{x}\mu_{\underline{0}}(u|0)\der u= 1-\pi_{0,\underline{x}}(0)$, yielding:
\begin{equation}\label{eq:max_before_FPT_1}
	\mu_{\underline{0}}(x|0)=-\frac{\der}{\der x}\pi_{0,\underline{x}}(0),
\end{equation} 
\noindent valid for general jump processes. As was recently shown in \cite{Klinger2022}, the splitting probability can only be computed explicitly for a handful of jump distributions; however, in the large $x$ limit, $\pi_{0,\underline{x}}(0)$ takes a universal asymptotic form which we readily exploit to obtain the large $x$ behavior of $\mu_{\underline{0}}(x|0)$:
\begin{equation}\label{eq:max_before_FPT_2}
	\mu_{\underline{0}}(x|0)\underset{x\to \infty}{\sim}\frac{\mu 2^{\mu-2}}{\sqrt{\pi}} \Gamma\left(\frac{1+\mu}{2}\right)\left[\frac{a_\mu}{x}\right]^{\frac{\mu}{2}}\frac{1}{x}.
\end{equation} 
\noindent Of note, the asymptotic decay is much slower than for fixed-length unbounded jump processes \eqref{eq:max_distrib_3}. Indeed, the survival probability $q(0,n)$ is decaying slowly enough to allow for particles to reach farther maxima before first crossing of 0. 

We now investigate joint space and time distributions. It is clear that being a solely geometrical quantity, $\pi_{0,\underline{x}}(0)$ is not sufficient to compute such joint distributions. In fact, in this case of bounded trajectories, $G_0(x,n)$ does not suffice to build EVS distributions.
To proceed further, we introduce the \textit{strip probability} $\mu_{0,\underline{x}}(n)$, defined as the probability that the particle starting from 0 stays positive and reaches its maximum $x$ on its $n^\text{th}$ step exactly, and show that $\mu_{0,\underline{x}}(n)$ allows for the systematic derivation of joint distributions.  Computing the exact expression of the strip probability requires two auxiliary quantities: (i) the joint distribution $\sigma(x,n_f|0)$ of the maximum $x$ and first passage time $n_f$ through 0 and (ii) the rightward exit time probability (RETP) $F_{0,\underline{x}}(n|x_0)$, defined as the probability that the particle crosses $x$ before 0 on its $n^{\text{th}}$ step exactly, which has been studied in \cite{Klinger:2022vb} (see SM for a summary of results). First, by partitioning trajectories over the time $k$  at which the maximum is reached, $\sigma$ is re-expressed in terms of $\mu_{0,\underline{x}}(n)$ and $F_{0,\underline{x}}(n|0)$ only:
\begin{equation}\label{eq:mu_1}
	\sigma(x,n|0)=\sum_{k=1}^{n-1}\mu_{0,\underline{x}}(k)F_{0,\underline{x}}(n-k|0).
\end{equation}
\noindent Next, we make use of the fact that the cumulative distribution of $\sigma$ is in fact given by $F_{\underline{0},x}(n|0)=\int_0^x \sigma(u,n|0)\der u$ where $F_{\underline{0},x}(n|x_0)=F_{0,\underline{x}}(n|x-x_0)$ by symmetry. Finally, we derive the exact expression of the generating function of the strip probability:
\begin{equation}\label{eq:mu_2}
	\sum_{n=1}^{\infty}\xi^n\mu_{0,\underline{x}}(n)=\frac{\frac{\der}{\der x} \tilde{F}_{\underline{0},x}(\xi|0)}{\tilde{F}_{0,\underline{x}}(\xi|0)}.
\end{equation}
\noindent Computing $\tilde{F}_{0,\underline{x}}(\xi|x_0)$ is thus sufficient to obtain explicit expressions of $\mu_{0,\underline{x}}(n)$. As an illustration, in the specific case of the exponential jump process we obtain
\begin{equation}\label{eq:mu_3}
	\tilde{\mu}_{0,\underline{x}}(\xi)=\frac{\text{sech}\left(\gamma  \sqrt{1-\xi } x\right)\gamma  (1-\xi) \xi }{(2-\xi) \sqrt{1-\xi } \tanh \left(\gamma  \sqrt{1-\xi } x\right)+2 (1-\xi) }.
\end{equation}
\noindent For general jump processes for which the RETP cannot be obtained explicitly, we analyze the large $x$ and $n$ behavior of $\mu_{0,\underline{x}}(n)$ and uncover emergent universal behavior.

In the $\mu=2$ case and in the scaling limit $\tau=n/n_x$ fixed, no overshoot occurs as the particle crosses $x$ for the first time. As a result, the events of crossing $x$ and reaching $x$ on the $n^{\text{th}}$ step become statistically equivalent, such that $\mu_{0,\underline{x}}(n)\sim a_2^{-1} F_{0,\underline{x}}(n|0)$, where the proportionality constant is fixed by using the exact exponential distribution result \eqref{eq:mu_3}. In turn, the asymptotic behavior of the strip probability is given by
\begin{equation}\label{eq:mu_5}
	\mu_{0,\underline{x}}(n) \underset{\substack{\tau \text{\ fixed}}}{\sim} 2\left[\frac{a_2}{x}\right]^{2}\frac{1}{x}\pi^2\sum_{k=1}^\infty k^2(-1)^{k+1}e^{-k^2\pi^2\tau}.
\end{equation}
\noindent For heavy-tailed jump processes, overshoots occur even in the limit $x\to\infty$, such that the identification of the strip probability and the RETP is no longer valid. Additionally, the exact expression \eqref{eq:mu_2} cannot be used to asymptotically analyze $\mu_{0,\underline{x}}(n)$; indeed $\tilde{F}_{\underline{0},x}(\xi|0)\sim 1/\sqrt{1-\xi}$ to leading $x$ order, so that $\frac{\der }{\der x}\tilde{F}_{\underline{0},x}(\xi|0)=0$. To circumvent these difficulties, we introduce the cumulative strip probability $\mu_{0,\underline{>x}}(n)=\int_{x}^{\infty}\mu_{0,\underline{u}}(n)\der u$, and partition trajectories over the step at which $x$ is crossed for the first time, and the ending position $u$ of the particle after the jump, whose probability distribution is denoted $\hat{F}_{0,\underline{x}}(u,k|0)$. In turn, the cumulative strip probability is written exactly as
\begin{equation}\label{eq:mu_levy_1}
		\hspace{-10pt}	\mu_{0,\underline{>x}}(n)=\sum_{k=1}^n \int_x^\infty \hat{F}_{0,\underline{x}}(u,k|0)\int_0^\infty G_{[0,y+u]}(y,n-k|0)\der y \der u,
\end{equation}
\noindent where $G_{[0,y]}(x,n|x_0)$ is the bounded propagator of a process killed upon first exit of the interval $[0,y]$. Importantly, in the large $x$ limit, since $u>x$, we have $G_{[0,y+u]}(y,n-k|0) \sim G_0(y,n-k|0)$, such that the cumulative strip probability is asymptotically given by
\begin{equation}\label{eq:mu_levy_2}
		\mu_{0,\underline{>x}}(n) \underset{n\ll n_x}{\sim} \sum_{k=1}^n F_{0,\underline{x}}(k|0)q(0,n-k),
\end{equation}
\noindent where we have used that $\int_{x}^{\infty} \hat{F}_{0,\underline{x}}(u,k|0)\der u=F_{0,\underline{x}}(k|0)$. Finally, we extract from equation \eqref{eq:mu_levy_2} the asymptotic universal behavior of $\mu_{0,\underline{x}}(n)$: 

\begin{equation}\label{eq:mu_levy_3}
	\mu_{0,\underline{x}}(n)\underset{1\ll n\ll n_x}{\sim} \frac{\mu}{\pi}\ \Gamma(\mu)\sin\left(\frac{\pi\mu}{2}\right)\left[\frac{a_\mu}{x}\right]^\mu \frac{1}{x}.
\end{equation}

\noindent Remarkably, $\mu_{0,\underline{x}}(n)$ becomes independent of $n$, in striking contrast with its unbounded counterpart $G_0(x,n)$.
Note also that, surprisingly, $\mu_{0,\underline{x}}(n)\sim p(x)$.
We now show that distributions of EVS observables for semi-infinite jump processes can be systematically obtained from the strip probability, and exploit the asymptotic results \eqref{eq:mu_5} and \eqref{eq:mu_levy_3} to derive explicit universal formulas.

As a first illustration, we determine the joint distribution $\rho_3(x,n_m|0)$ of the maximum and time at which it is reached. Paralleling the unbounded result \eqref{eq:rho1_1}, we decompose the Markovian trajectory into two independent parts around $n_m$ (see figure \ref{fig:schem}(b)), and identify their respective probabilistic weights to obtain

\begin{equation}\label{eq:rho3_1}
	\rho_3(x,n_m|0)= \mu_{0,\underline{x}}(n_m)\pi_{0,\underline{x}}(0).
\end{equation}

\noindent 
Making use of the asymptotic behavior of the strip probability given above, we derive large $x$ and $n$ expressions of $\rho_3$. For $\mu=2$, the joint distribution reads

\begin{equation}\label{eq:rho3_2}
	\rho_3(x,n_m|0)\underset{\substack{ \tau \text{\ fixed}}}{\sim} 2\left[\frac{a_2}{x}\right]^{3}\frac{1}{x}\pi^2\sum_{k=1}^\infty k^2(-1)^{k+1}e^{-k^2\pi^2\tau},
\end{equation}

\noindent while in the heavy-tailed case one has

\begin{equation}\label{eq:rho3_3}
	\rho_3(x,n|0)\underset{\substack{1\ll n\ll n_x}}{\sim}\frac{2^{\mu-1}\ \mu\ \Gamma(\frac{1+\mu}{2})\ \Gamma(\mu)\ \sin(\frac{\pi\mu}{2})}{\pi^{\frac{3}{2}}}\left[\frac{a_\mu}{x}\right]^{\frac{3\mu}{2}}\frac{1}{x}.
\end{equation}

\noindent Importantly, the $n$-independence of the strip probability has drastic effects on $\rho_3$; indeed, conditioned on the value $x$ of the maximum, the time at which it is reached becomes equiprobable for values of $n \ll n_x$.

As a second illustration,
we obtain thanks to this formalism the joint distribution $\rho_4(x,n_m,n_f|0)$ of the maximum $x$, time of maximum $n_m$ and first passage time $n_f$ across 0, which is given by:
\begin{equation}\label{eq:rho_4}
\rho_4(x,n,n_f|0)=\mu_{0,\underline{x}}(n)F_{0,\underline{x}}(n_f-n).
\end{equation}  
\noindent Finally, its asymptotic behavior is readily obtained from that of the strip probability, and we provide universal formulas in SM, along with the analysis of the joint distribution $\sigma(x,n_f|0)$ of the maximum and first passage time across 0.

\textit{Conclusion} We have shown that for general symmetric jump processes, the derivation of joint space and time distributions of EVS observables reduces to the determination of a single key quantity, which only depends on the geometrical constraints imposed on the trajectory.
For unbounded jump processes, we identified the sufficient building block to be the semi-infinite propagator $G_0(x,n)$ and made use of its $\mu$-dependent limit behavior to draw a comprehensive picture of large space and time EVS asymptotics. In the case of semi-infinite jump processes killed upon first crossing of 0, $G_0(x,n)$ is ill-fitted to investigate EVS observables. As a replacement,  we introduced the strip probability $\mu_{0,\underline{x}}(n)$, provided exact and asymptotic expressions valid for general symmetric jump distribution $p(\ell)$, and systematically derived joint EVS distributions summarized in table \ref{tab:summary}. In addition to these asymptotic results, we emphasize that all distributions can be explicitly computed for any $n$ and $x$ values, as soon as $G_0(x,n)$ and $\mu_{0,\underline{x}}(n)$ are known, as is the case for the exponential jump process $p(\ell)=\frac{\gamma}{2}e^{-\gamma |\ell|}$, a paradigmatic model of single active particle motion.


\begin{table}[h!]
			\begin{tabular}{|cccc|}
				\hline
				
				\multicolumn{2}{|c}{Unbounded jump processes } & \multicolumn{2}{||c|}{Semi-Infinite jump processes} \\ \hline \hline
				
				\multicolumn{1}{|c}{$\mu(x|n)$ \cite{Bingham:1973, Majumdar:2010r}} & 	\multicolumn{1}{|c}{$\large \color{red}\checkmark$} & 	\multicolumn{1}{||c|}{$\mu_{\underline{0}}(x|x_0=0)$} & $\large \color{green} \checkmark $ \\ \hline
				
				\multicolumn{1}{|c}{$\rho(n_m|n)$ \cite{Majumdar:2010r}} & 	\multicolumn{1}{|c}{$\large \color{red} \checkmark$} & 	\multicolumn{1}{||c|}{$\rho_3(x,n_m|x_0=0)$} & $\large \color{green} \checkmark$ \\ \hline
				
				\multicolumn{1}{|c}{$\rho_1(x,n_m|n)$} & 	\multicolumn{1}{|c}{$\large \color{green} \checkmark$} & 	\multicolumn{1}{||c|}{$\rho_4(x,n_m,n_f|x_0=0)$} &$\large \color{green} \checkmark$ \\ \hline
				
				\multicolumn{1}{|c}{$\rho_2(x,n_m,x_f|n)$} & 	\multicolumn{1}{|c}{$\large \color{green} \checkmark$} & 	\multicolumn{1}{||c|}{$\sigma(x,n_f|x_0=0)$} & $\large \color{green} \checkmark$ \\ \hline
				
			\end{tabular}
	\caption{EVS observables for general unbounded and semi-infinite jump processes. The variables $x,n_m, x_f$ and $n_f$ respectively denote the maximum, the time at which the maximum is reached, the final position of the process and the first passage time across 0. Our framework allows for the computation of novel exact and asymptotic expressions for all distributions labeled with $ \color{green} \checkmark$ - explicit expressions are given in Table I of the SM.  Entries labeled with $ \color{red} \checkmark$ are already given in the literature.}
	\label{tab:summary}
	
\end{table}

%
%
%
%
%
%
%
%
%
%
%
%

\bibliographystyle{apsrev4-2}
%



\section*{Supplementary material}

# Supplementary Material for "Extreme Value Statistics of Jump Processes"

J. Klinger,[1] R. Voituriez,[2] and O. Bénichou[1]

[1]*Laboratoire de Physique Théorique de la Matière Condensée,*
*CNRS, UPMC, 4 Place Jussieu, 75005 Paris, France*
[2]*Laboratoire Jean Perrin, CNRS, UPMC, 4 Place Jussieu, 75005 Paris, France*

This Supplementary Material presents technical details:

- On explicit expressions of EVS distributions

- On the running maximum of stable processes

- On the asymptotic behavior of $q(x, n)$ and $G_0(x, n)$

- On the RETP and LETP

- On non zero initial conditions

- On lattice random walks

## I. ON EXPLICIT EXPRESSIONS OF EVS DISTRIBUTIONS

In this section, we provide a table summarizing the exact and asymptotic expression of the various EVS joint distributions discussed in the main text. Exact expressions mostly stem from Markovian decomposition of trajectories, while the asymptotics are obtained from the analysis of the two essential building blocks $G_0(x, n)$ and the strip probability $\mu_{0,x}(n)$. Note that in the specific case of $\rho_1$ and $\rho_2$, the asymptotic behavior for processes with $\mu = 2$ is simply given by the limit result for Brownian motion with diffusion coefficient $D$ and duration $t$, which can be found in [2] for instance. In practice, this amounts to equating $n = t$ and $D = a_2^2$.

| EVS observable | Exact expression | $\mu < 2$ - Asymptotic behavior in the regime $1 \ll n \ll n_x$ | $\mu = 2$ - Asymptotic behavior in the scaling regime $n/n_x$ fixed |
|---|---|---|---|
| | | **Unbounded jump processes** | |
| $\mu(x\|n)$ | $\frac{\mathrm{d}}{\mathrm{d}x} q(x, n)$ | $\frac{n\mu}{\pi} \sin\left(\frac{\pi\mu}{2}\right) \Gamma(\mu) \left[\frac{a_\mu}{x}\right]^\mu \frac{1}{x}$  o | $\frac{1}{a_2\sqrt{\pi n}} e^{-\left[\frac{x}{a_2}\right]^2 \frac{1}{4n}}$  o |
| $\rho(n_m\|n)$ | $q(0, n_m) q(0, n - n_m)$ | $\frac{1}{\pi} \frac{1}{\sqrt{n_m(n - n_m)}}$  o | |
| $\rho_1(x, n_m\|n)$ | $G_0(x, n_m) q(0, n - n_m)$ | $\frac{1}{\pi} \sqrt{\frac{n_m}{n - n_m}} \frac{2\mu}{x} \sin\left(\frac{\pi\mu}{2}\right) \Gamma(\mu) \left[\frac{a_\mu}{x}\right]^\mu \frac{1}{x}$ | $\frac{x}{a_2} \frac{1}{2a_2\pi n_m^{3/2}\sqrt{n - n_m}} e^{-\left[\frac{x}{a_2}\right]^2 \frac{1}{4n_m}}$ |
| $\rho_2(x, n_m, x_f\|n)$ | $G_0(x, n_m) G_0(x - x_f, n - n_m)$ | $\sqrt{n_m(n - n_m)} \frac{4\mu^2}{\pi^2} \sin^2\left(\frac{\pi\mu}{2}\right) \Gamma^2(\mu) \left[\frac{a_\mu^2}{x(x - x_f)}\right]^\mu \frac{1}{x(x - x_f)}$ | $\frac{x}{a_2} \frac{x - x_f}{a_2} \frac{1}{4(a_2)^2 \pi(n_m(n - n_m))^{3/2}} e^{-\left[\frac{x}{a_2}\right]^2 \frac{1}{4n_m} - \left[\frac{x - x_f}{a_2}\right]^2 \frac{1}{4(n - n_m)}}$ |
| | | **Semi-infinite jump processes** | |
| $\mu_0(x\|x_0 = 0)$ | $-\frac{\mathrm{d}}{\mathrm{d}x} \pi_{0,x}(0)$ | | $\mu 2^{\mu - 2} \Gamma\left(\frac{1 + \mu}{2}\right) \frac{1}{\sqrt{\pi}} \left[\frac{a_\mu}{x}\right]^{\frac{\mu}{2}} \frac{1}{x}$ |
| $\rho_3(x, n_m\|x_0 = 0)$ | $\mu_{0,x}(n_m) \pi_{0,x}(0)$ | $\frac{2^{\mu - 1}}{\pi^{\frac{3}{2}}} \frac{\mu \; \Gamma\left(\frac{1 + \mu}{2}\right) \; \Gamma(\mu) \; \sin\left(\frac{\pi\mu}{2}\right)}{\pi^{\frac{3}{2}}} \left[\frac{a_\mu}{x}\right]^{\frac{3\mu}{2}} \frac{1}{x}$ | $2 \left[\frac{a_2}{x}\right]^3 \frac{1}{x} \pi^2 \sum_{k=1}^\infty k^2(-1)^{k+1} e^{-k^2\pi^2 n_m \left[\frac{a_2}{x}\right]^2}$ |
| $\rho_4(x, n_m, n_f\|x_0 = 0)$ | $\mu_{0,x}(n_m) F_{0,x}(n_f - n_m\|0)$ | $\frac{\mu}{\sqrt{\pi(n_f - n)}} \left[\frac{\Gamma(\mu)}{\pi} \sin\left(\frac{\pi\mu}{2}\right)\right]^2 \left[\frac{a_\mu}{x}\right]^{2\mu} \frac{1}{x}$ | $4 \left[\frac{a_2}{x}\right]^5 \frac{1}{x} \pi^4 \sum_{k,l=1}^\infty k^2 l^2(-1)^{k+l} e^{-k^2\pi^2 n_m \left[\frac{a_2}{x}\right]^2 - l^2\pi^2(n_f - n_m)\left[\frac{a_2}{x}\right]^2}$ |
| $\sigma(x, n_f\|x_0 = 0)$ | $\frac{\mathrm{d}}{\mathrm{d}x} F_{0,x}(n_f\|0)$ | $\frac{2\mu\sqrt{n_f}}{\pi^{\frac{3}{2}}} \Gamma^2(\mu) \sin^2\left(\frac{\pi\mu}{2}\right) \left[\frac{a_\mu}{x}\right]^{2\mu} \frac{1}{x}$ | $2\pi^2 \left[\frac{a_2}{x}\right]^3 \frac{1}{x} \sum_{k=1}^\infty e^{-k^2\pi^2 n_f \left[\frac{a_2}{x}\right]^2} k^2 \left[2k^2\pi^2 n_f \left[\frac{a_2}{x}\right]^2 - 3\right]$ |

TABLE I. EVS observables for general jump processes. For the asymptotic behavior, $n_x$ is defined as $n_x = (x/a_\mu)^\mu$ for all $\mu$ values, and $x$, $n_m$ and $n_f$ correspond respectively to the maximum, time at which the maximum is reached, and first-passage time across 0. Entries with a o are already given in the literature.

## II. ON THE RUNNING MAXIMUM OF STABLE PROCESSES

### A. Exact expressions and two explicit cases

In this section, we list useful results from Darling [4] and Bingham [1] on the distribution of the running maximum of stable processes. Although their results extend to non-symmetric stable processes, we focus here on strictly symmetric processes $X(t)$, defined by their characteristic function

$$\mathbb{E}[e^{isX(t)}] = \exp\left[-ct|s|^{\mu}\right], \tag{S1}$$

where $\mu$ is the Levy index of the stable process and $c$ some positive constant. Denoting now the running maximum $M(t)$, the scaling properties of stable processes imply that $M(t)/t^{\frac{1}{\mu}}$ is distributed as $M(1)$, whose Laplace transform is denoted

$$\mathbb{E}[e^{-sM(1)}] = \phi(s). \tag{S2}$$

Importantly, it was shown by Darling that $\phi(s)$ takes the following form

$$\int_0^\infty \phi(sx^{\frac{1}{\mu}})\mathrm{d}x = g(s)$$
$$g(s) = \exp\left[-\frac{1}{\pi}\int_0^\infty \frac{\ln\left(1 + c(sx)^\mu\right)}{x^2 + 1}\mathrm{d}x\right]. \tag{S3}$$

While it is difficult to obtain explicit expressions for the distribution $f(x)$ of $M(1)$ from equation (S3), there exists two exact results:

- In the Brownian case $\mu = 2$, one has

$$f^B(x) = \frac{\exp\left[-x^2/(4c)\right]}{\sqrt{\pi c}}. \tag{S4}$$

- In the Cauchy case $\mu = 1$

$$f^C(x) = \frac{f^*(\frac{x}{c})}{c}; \quad f^*(x) = \frac{1}{\pi\sqrt{x(1 + x^2)^{\frac{3}{2}}}}\exp\left[-\frac{1}{\pi}\int_0^x \frac{\ln(y)}{1 + y^2}\mathrm{d}y\right] \tag{S5}$$

Finally, In the case where the exact distribution of $M(1)$ cannot be explicitly computed, Bingham asymptotically studied equation (S3) to obtain the tail behavior of $M(1)$:

$$\mathbb{P}(M(1) > x) \underset{x\to\infty}{\sim} \frac{c\Gamma(\mu)\sin\left(\frac{\pi\mu}{2}\right)}{\pi x^\mu}. \tag{S6}$$

### B. Scaling limit for jump processes with $\mu < 2$

To identify the scaling limit distribution of the running maxima $M_n$ of a heavy-tailed jump process characterized by $\mu$ and $a_\mu$, one has to identify the limit $\alpha$-stable process. In the case of symmetric jump distributions, this is straightforward (see [3] for a review) and amounts to equating $n = t$, as well as $c = [a_\mu]^\mu$. In turn, by making use of equation (S6) as well as the scale invariance of $M(t)$, the algebraic decay of the distribution of $M_n$ (equation (7) from the main text) is recovered.

## III.   ON THE ASYMPTOTIC BEHAVIOR OF $q(x,n)$ AND $G_0(x,n)$

In this section, we provide details on the asymptotic analysis of $q(x,n)$ and $G_0(x,n)$ in the $1 \ll n \ll n_x$ regime. We focus on heavy-tailed processes with $\mu < 1$ to lighten calculations, but the results are valid for all $\mu < 2$.

### A.   Semi-infinite propagator

We start from the Laplace transform of $G_0(x,n)$:

$$\tilde{G}_0(s,\xi) \equiv \sum_{n=1}^{\infty} \xi^n \left[ \int_0^{\infty} e^{-sx} G_0(x,n) \mathrm{d}x \right] = \exp\left[ -\frac{s}{2\pi} \int_{-\infty}^{\infty} \frac{\ln\left[1 - \xi \tilde{p}(k)\right]}{s^2 + k^2} \mathrm{d}k \right], \tag{S7}$$

and study the small $s$ limit. Making a change of variables yields

$$\tilde{G}_0(s,\xi) = \exp\left[ -\frac{1}{\pi} \int_0^{\infty} \frac{\ln\left[1 - \xi \tilde{p}(sk)\right]}{1 + k^2} \mathrm{d}k \right]. \tag{S8}$$

In the small $s$ limit, one has $\tilde{p}(ks) = 1 - (a_\mu s|k|)^\mu + o(|k|^\mu)$, and the above expression reads

$$\tilde{G}_0(s,\xi) \sim \exp\left[ -\ln(1 - \xi)\frac{1}{2} \right] \exp\left[ \frac{(a_\mu s)^\mu}{(1 - \xi)\pi} \int_0^{\infty} \frac{k^\mu}{1 + k^2} \mathrm{d}k \right] \tag{S9}$$

and to first vanishing order in $s$:

$$\tilde{G}_0(s,\xi) \sim \frac{1}{\sqrt{1 - \xi}} \left[ 1 + \frac{(a_\mu s)^\mu}{(1 - \xi)\pi} \int_0^{\infty} \frac{k^\mu}{1 + k^2} \mathrm{d}k \right] + o(s^\mu). \tag{S10}$$

The leading order is given by

$$\tilde{G}_0(s,\xi) = \frac{1}{\sqrt{1 - \xi}} = \sum_{n=1}^{\infty} \xi^n q(0,n), \tag{S11}$$

and we deduce from the first vanishing order the algebraic decay of $G_0(x,n)$ in the $1 \ll n \ll n_x$ regime:

$$G_0(x,n) \underset{1 \ll n \ll n_x}{\sim} 2\sqrt{\frac{n}{\pi}} \frac{\mu}{\pi} \sin\left(\frac{\pi\mu}{2}\right) \Gamma(\mu) \left[\frac{a_\mu}{x}\right]^\mu \frac{1}{x}, \tag{S12}$$

which is used to obtain asymptotic expressions for $\rho_1$ and $\rho_2$ in the main text.

### B.   Survival probability

We proceed similarly to analyze the large $x$ decay of $q(x,n)$, the survival probability starting from $x$. We start from the Laplace transform of the survival probability, obtained from equation (4) of the main text:

$$\sum_{n=0}^{\infty} \xi^n \left[ \int_0^{\infty} e^{-sx_0} q(x_0,n) \mathrm{d}x_0 \right] = \frac{1}{s\sqrt{1 - \xi}} \exp\left[ -\frac{s}{2\pi} \int_{-\infty}^{\infty} \frac{\ln\left[1 - \xi \tilde{p}(k)\right]}{s^2 + k^2} \mathrm{d}k \right]$$
$$= \frac{1}{s\sqrt{1 - \xi}} \tilde{G}_0(s,\xi). \tag{S13}$$

By making use of the expansion (S12), we find the leading vanishing order of $q(x,n)$:

$$1 - q(x,n) \underset{1 \ll n \ll n_x}{\sim} \frac{n}{\pi} \sin\left(\frac{\pi\mu}{2}\right) \Gamma(\mu) \left[\frac{a_\mu}{x}\right]^\mu, \tag{S14}$$

yielding the asymptotic behavior of distribution $\mu(x|n)$ of the running maximum for heavy tailed processes given in the main text (7).

## IV. ON THE RETP AND LETP

In this section, we briefly review results from [7] on the rightward exit time probability $F_{0,\underline{x}}(n|x_0)$ (RETP), defined as the probability that a jump process crosses $x$ for the first time on its $n^{\text{th}}$ step, without having crossed 0 before, as well as its leftward counterpart $F_{\underline{0},x}(n|x_0)$ (LETP). By partitioning over the first step of the walk, the RETP is shown to satisfy the following integral equation:

$$F_{0,\underline{x}}(n|x_0) = (1 - \delta_{n,1}) \int_0^x p(y - x_0) F_{0,\underline{x}}(n - 1|y) \mathrm{d}y + \delta_{n,1} \left[ \int_x^\infty p(y - x_0) \mathrm{d}y \right]. \tag{S15}$$

For arbitrary jump kernels $p(\ell)$, such integral equations are not exactly solvable. However in the specific case of the exponential jump process $p(\ell) = \frac{1}{2} e^{-|\ell|}$, the RETP is explicitly given in the generating function formalism:

$$\sum_{n=1}^\infty F_{0,\underline{x}}(n|x_0)\xi^n = \frac{\xi \left( \sqrt{1 - \xi} \cosh(x_0\sqrt{1 - \xi}) + \sinh(x_0\sqrt{1 - \xi}) \right)}{2\sqrt{1 - \xi} \cosh(x\sqrt{1 - \xi}) - (\xi - 2) \sinh(x\sqrt{1 - \xi})}, \tag{S16}$$

from which, using symmetry arguments, we obtain the explicit strip probability for the exponential jump process - see equation (16) in the main text. For general jump processes, the asymptotic behavior of the RETP can be studied in the scaling limit $\tau = \frac{n}{n_x}$ fixed, with $n_x = [x/a_\mu]^\mu$, and known results are summarized in the following table

| | $F_{0,\underline{x}}(n|x_0) \underset{\substack{n \to \infty \\ x \to \infty \\ \tau \text{ fixed}}}{\sim} \pi_{0,\underline{x}}(x_0)h_\mu(\tau)n^{-1}$ | $F_{\underline{0},x}(n|x_0) \underset{\substack{n \to \infty \\ x \to \infty \\ \tau \text{ fixed}}}{\sim} F_{\underline{0}}(n|x_0)g_\mu(\tau)$ |
|---|---|---|
| $\mu = 2$ | $h_2(\tau) = 2\pi\tau^2 \sum_{k=1}^\infty k^2(-1)^{k+1} e^{-k^2\pi^2\tau}$ | $g_2(\tau) = 4\pi^{\frac{5}{2}}\tau^{\frac{3}{2}} \sum_{k=1}^\infty e^{-k^2\pi^2\tau} k^2$ |
| $\mu < 2$ | $h_\mu(\tau) \underset{\tau \ll 1}{\sim} \Gamma(\mu/2) \sin(\pi\mu/2)\pi^{-\frac{3}{2}}\sqrt{\tau}$ | $g_\mu(\tau) \underset{\tau \ll 1}{\sim} 1$ |
| | $h_\mu(\tau) \underset{\tau \gg 1}{\sim} \gamma_\mu \left[ \lambda_1 2^\mu \tau \right] e^{-\lambda_1 2^\mu \tau}$ | $g_\mu(\tau) \underset{\tau \gg 1}{\sim} \omega_\mu \left[ \lambda_1 2^\mu \tau \right]^{3/2} e^{-\lambda_1 2^\mu \tau}$ |

TABLE II. Asymptotic behavior of RETP and RETP.

where $F_{\underline{0}}(n|x_0)$ is the distribution of the first crossing time of zero for a semi-infinite jump process, and the constants

$$\begin{aligned}
\gamma_\mu &\equiv \sqrt{\mu} 2^{\frac{\mu}{2} - \frac{3}{2}} \frac{\Gamma\left(\frac{\mu}{2}\right)}{\Gamma(\mu)} \left[ \frac{\pi}{2} - \frac{(2 - \mu)\pi}{8} \right]^{\frac{\mu}{2}} \int_0^2 \psi_1(u)\mathrm{d}u \\
\omega_\mu &\equiv \sqrt{\frac{2\pi}{\mu}} \frac{\Gamma\left(1 + \frac{\mu}{2}\right)}{\Gamma\left(\frac{\mu}{2}\right)} \int_0^2 \psi_1(u)\mathrm{d}u
\end{aligned} \tag{S17}$$

are given in terms of the first eigenfunction $\psi_1(u)$ of the fractional diffusion equation on $[0, 2]$ (see [8] for approximate expressions).

### A. Asymptotic Strip probability for heavy-tailed processes

Importantly, it can be seen from table II that the RETP for heavy-tailed processes in the regime $1 \ll n \ll n_x$ can be rewritten in the following form

$$F_{0,\underline{x}}(n|0) \underset{x \to \infty}{\sim} q(0, n - 1) \frac{\Gamma(\mu)}{\pi} \sin\left(\frac{\pi\mu}{2}\right) \left[ \frac{a_\mu}{x} \right]^\mu, \tag{S18}$$

where the survival probability $q(0, n)$ for semi-infinite processes starting from zero is given in the generating function formalism by

$$\sum_{n=0}^\infty \xi^n q(0, n) = \frac{1}{\sqrt{1 - \xi}}. \tag{S19}$$

Equations (S18) and (S19) taken together yield the asymptotic behavior of the strip probability for heavy-tailed processes, given in equation (20) of the main text.

### B. $\sigma(x, n_f | 0)$ in the case $\mu = 2$

We show that for processes with $\mu < 2$, the joint distribution $\sigma(x, n_f | 0)$ of the maximum $x$ and first crossing time $n_f$ of 0 is obtained from the LETP $F_{\underline{0}, x}(n | 0)$. By partitioning over the maximum value reached by the particle before crossing 0, it is clear that

$$F_{\underline{0}, x}(n | 0) = \int_0^x \sigma(u, n | 0) \mathrm{d}u, \tag{S20}$$

or equivalently

$$\sigma(x, n | 0) = \frac{\mathrm{d}}{\mathrm{d}x} F_{\underline{0}, x}(n | 0). \tag{S21}$$

In turn, in the scaling limit $n \gg 1$, $x \gg 1$ and $\tau = n(a_2 / x)^2$ fixed, the LETP is given (see table II) by

$$F_{\underline{0}, x}(n | 0) \underset{\tau \text{ fixed}}{\sim} \frac{1}{2\sqrt{\pi n^3}} 4\pi^{\frac{5}{2}} \tau^{\frac{3}{2}} \sum_{k=1}^{\infty} e^{-k^2 \pi^2 \tau} k^2 \tag{S22}$$

such that we obtain

$$\begin{aligned}
\sigma(x, n | 0) &\underset{\tau \text{ fixed}}{\sim} -\frac{1}{2\sqrt{\pi n^3}} 4\pi^{\frac{5}{2}} \tau^{\frac{1}{2}} \sum_{k=1}^{\infty} e^{-k^2 \pi^2 \tau} k^2 \left[\frac{3}{2} - k^2 \pi^2 \tau\right] \times \tau \frac{2}{x} \\
&\underset{\tau \text{ fixed}}{\sim} -\frac{1}{\sqrt{n^3}} 2\pi^2 \tau^{\frac{3}{2}} \sum_{k=1}^{\infty} e^{-k^2 \pi^2 \tau} k^2 \left[3 - 2k^2 \pi^2 \tau\right] \times \frac{1}{x} \\
&\underset{\tau \text{ fixed}}{\sim} 2\pi^2 \left[\frac{a_2}{x}\right]^3 \frac{1}{x} \sum_{k=1}^{\infty} e^{-k^2 \pi^2 \tau} k^2 \left[2k^2 \pi^2 \tau - 3\right]
\end{aligned} \tag{S23}$$

which is identical to the result in table I.

## V. ON THE NON ZERO INITIAL CONDITION

In this section, we discuss the impact of non zero initial conditions on EVS of jump processes.

### A. Unbounded jump processes

In the case of unbounded jump processes, non zero initial conditions are straightforwardly dealt with. Indeed, all distributions $\mu$, $\rho$, $\rho_1$ and $\rho_2$ are translation invariant, such that for a generic distribution $\tilde{\rho}(x | n, x_0)$, one has

$$\tilde{\rho}(x | n, x_0) = \tilde{\rho}(x - x_0 | n), \tag{S24}$$

which is valid for all $x_0$ initial conditions.

## B.  Semi-infinite jump processes

For semi-infinite jump processes killed upon first crossing of 0, we show that our framework captures the specific small $x_0$ behavior arising from the discrete nature of jump processes. Consider first the exact expressions given in table I. By using the Markovian nature of jump process trajectories, is is easily seen that the $x_0$ dependence only enters through the first part of the trajectory decomposition such that:

$$
\begin{aligned}
\mu_{\underline{0}}(x|x_0) &= -\frac{\mathrm{d}}{\mathrm{d}x}\pi_{0,\underline{x}}(x_0), \\
\rho_3(x, n_m|x_0) &= \mu_{0,\underline{x}}(n_m|x_0)\pi_{0,\underline{x}}(0), \\
\rho_4(x, n_m, n_f|x_0) &= \mu_{0,\underline{x}}(n_m|x_0)F_{0,\underline{x}}(n_f-n_m|0), \\
\sigma(x, n_f|x_0) &= \frac{\mathrm{d}}{\mathrm{d}x}F_{\underline{0},x}(n_f|x_0).
\end{aligned}
\tag{S25}
$$

Note that the strip probability $\mu_{0,\underline{x}}(n|x_0)$ is now defined as the probability that the particle originating from $x_0 > 0$ stays positive during its first $n$ steps, and reaches its maximum $x$ on its last step exactly. The $x_0$ dependence of both $\pi_{0,\underline{x}}(x_0)$ and the LETP $F_{\underline{0},x}(n_f|x_0)$ have been extensively discussed in [6] and [7]; in particular, it was shown that in the large $x$ limit, the splitting probability is given by

$$
\begin{aligned}
\lim_{x\to\infty}\left[\frac{\pi_{0,\underline{x}}(x_0)}{A_\mu(x)}\right] &= \frac{1}{\sqrt{\pi}} + V(x_0) \\
A_\mu(x) &= \left(\frac{a_\mu}{x}\right)^{\mu/2}2^{\mu-1}\Gamma\left(\frac{1+\mu}{2}\right).
\end{aligned}
\tag{S26}
$$

where $V(x_0)$ is an explicit process-dependent function, given by its Laplace transform

$$
\int_0^\infty e^{-sx_0}V(x_0)\mathrm{d}x_0 = \frac{1}{s\sqrt{\pi}}\left(\exp\left[-\frac{s}{\pi}\int_0^\infty\frac{\mathrm{d}k}{s^2+k^2}\ln(1-\tilde{p}(k))\right]-1\right).
\tag{S27}
$$

Consequently, we only need to determine the $x_0$ dependence of the strip probability $\mu_{0,\underline{x}}(n|x_0)$ to obtain a comprehensive picture of EVS of semi-infinite jump processes. Our results hold in the general asymptotic framework discussed in the main text, namely the scaling regime for processes with $\mu = 2$, and the regime $1 \ll n \ll n_x$ for heavy-tailed processes. First, recall that

$$
\begin{aligned}
\mu = 2: &\qquad \mu_{0,\underline{x}}(n) \sim a_2^{-1}F_{0,\underline{x}}(n), \\
\mu < 2: &\qquad \mu_{0,\geq\underline{x}}(n) \underset{1\ll n\ll n_x}{\sim} \sum_{k=1}^n F_{0,\underline{x}}(k|0)q(0, n-k).
\end{aligned}
\tag{S28}
$$

In turn, we find that the $x_0$ dependence only enters through the RETP. Next, it can be seen from table II that $F_{0,\underline{x}}(n|x_0)/\pi_{0,\underline{x}}(x_0)$ is independent of $x_0$, such that the $x_0$ dependence of the two quantities is equal. As a result, we obtain the following large $x$ and $n$ asymptotic behavior

$$
\mu_{0,\underline{x}}(n|x_0) \sim \mu_{0,\underline{x}}(n)\left[1+\sqrt{\pi}V(x_0)\right],
\tag{S29}
$$

valid for all $x_0$ initial conditions, in particular the regime $x_0 \lesssim a_\mu$, for which discrete effects become crucial. As an illustration, we display on figure 1 numerical simulations accounting for the $x_0$ dependence of the cumulative strip probability $\mu_{0,\geq\underline{x}}(n|x_0)$.

## VI.  ON LATTICE RANDOM WALKS

In this section, we discuss the natural extension to Markovian lattice random walk, such as the one-dimensional Polya walk. In essence, all decompositions summarized in table I remain exact, although care needs to be taken in the

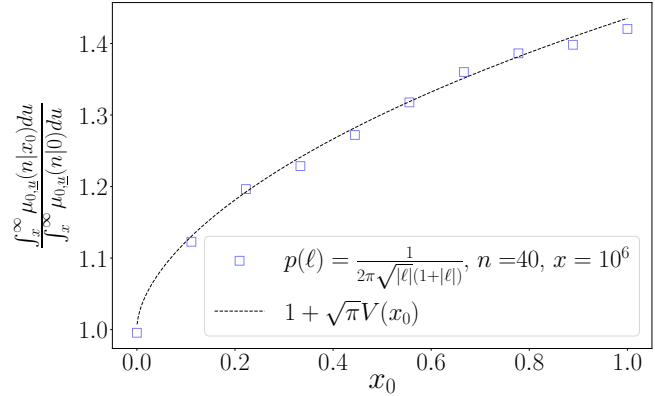

FIG. 1. $x_0$ dependence of the cumulative strip probability for a F-distributed heavy-tailed jump process with fixed $n$ and and $x$. The dashed black line is obtained from the numerical evaluation of $V(x_0)$ given in equation (S27).

definition of EVS joint distributions. Indeed, since lattice walks can repeatedly visit sites, it becomes necessary to distinguish trajectories allowed to come back to their starting point, from those that never visit their starting point twice. Importantly:

- We distinguish the conditional semi-infinite propagator $G_0^\dagger(s, n)$, defined as the probability that the walker originated from $s_0 = 0$ is at site $s$ on its $n^{\text{th}}$ step , and its successive positions $(s_1, \ldots, s_n)$ are *strictly* positive, from $G_0(s, n)$, where the walker is allowed to revisit 0.

- We distinguish the conditional survival probability $q^\dagger(0, n)$, defined as the probability that the walker originating from 0 has its successive positions $(s_1, \ldots, s_n)$ *strictly* positive, from $q(0, n)$, where the walker is allowed to revisit 0.

Joint EVS distributions of lattice walks can now be expressed in terms of these new quantities. As an example, in the case of unbounded lattice walks we obtain:

$$\rho(n_m|n) = q^\dagger(0, n)q(0, n - n_m)$$
$$\rho_1(s, n_m|n) = G_0^\dagger(s, n)q(0, n - n_m) \qquad \text{(S30)}$$
$$\rho_2(s, n_m, s_f|n) = G_0^\dagger(s, n)G_0(s - s_f, n - n_m)$$

where $n_m$ is now the *first* time the maximum $s$ is reached. For semi-infinite lattice walks terminated upon becoming strictly negative, no such distinction is needed and trajectories contributing to $\pi_{0,s}(0)$, $F_{0,s}(n)$ and $\mu_{0,s}(n)$ are allowed to return repeatedly to 0. Finally, we point out that in the case of the one dimensional Polya walk, explicit expressions for $G_0^\dagger(s, n)$, $G_0(s, n)$, $q^\dagger(0, n)$ and $q(0, n)$ can be found in [5].

---



\end{document}